\documentclass[a4paper,aps,prd,12pt,preprintnumbers,showpacs,onecolumn,superscriptaddress,nofootinbib,amsmath,amssymb]{revtex4-1}
\usepackage[dvips]{graphics}
\usepackage{cmap}
\usepackage[utf8]{inputenc}
\usepackage[T1]{fontenc}

\def\imo{i}
\def\re#1{Re(#1)}
\def\im#1{Im(#1)}
\def\K{{\cal K}}
\def\Order#1{{\cal O}\left(#1\right)}

\begin{document}
\title{Perturbations and Quasinormal Modes of the Dirac Field in Effective Quantum Gravity}
\author{Zainab Malik}\email{zainabmalik8115@outlook.com}
\affiliation{Institute of Applied Sciences and Intelligent Systems, H-15, Pakistan}
\begin{abstract}
Recently, two models of quantum-corrected Schwarzschild-like black holes were developed within Effective Quantum Gravity, and the spectra of bosonic perturbations have been analyzed in several recent studies. In this work, we investigate the quasinormal modes of a massless Dirac field perturbations around these black holes. By employing the higher-order WKB method and time-domain integration, we achieved consistency between the two approaches within their common range of validity. This concordance enables us to distinguish the two quantum-corrected black hole models from each other and from their Schwarzschild limit through their quasinormal spectra. In addition, we find approximate analytic expressions for quasinormal modes and grey-body factors in the form of expansion beyond the eikonal limit.
\end{abstract}
\maketitle
\section{Introduction}

Quasinormal modes (QNMs)  a black hole background describe the characteristic oscillations of  perturbations in the spacetime surrounding the black hole \cite{Nollert:1999ji,Kokkotas:1999bd,Konoplya:2011qq}. These modes are defined by complex frequencies, where the real part corresponds to the oscillation frequency, and the imaginary part is related to the damping rate due to the emission of radiation. Quasinormal modes arising from spacetime perturbations are observed through gravitational-wave interferometers~\cite{LIGOScientific:2016aoc,LIGOScientific:2017vwq,LIGOScientific:2018mvr}. Combined with observations in the electromagnetic spectrum \cite{EventHorizonTelescope:2019dse,EventHorizonTelescope:2019ggy,Goddi:2016qax}, these detections enable the study of gravity in the strong-field regime.

While the majority of publications on quasinormal modes are devoted to bosons fields, the fermionic perturbations play its role as cosmological background or astrophysical environment of a black hole. The Dirac field in this context is usually ascribed to neutrinos.  Properties of neutrinos, in its turn, are intrinsically bound to a number of  cosmological problems \cite{Lesgourgues:2006nd,Boyarsky:2009ix,Vagnozzi:2017ovm,Vagnozzi:2018jhn}. The study of Dirac QNMs is crucial for understanding the stability of  fermionic perturbations and provides insights into the behavior of quantum fields in curved spacetime. In the eikonal limit, the QNMs are closely related to the properties of null geodesics, particularly the orbital frequency and the Lyapunov exponent, which characterize the stability of these geodesics \cite{Cardoso:2008bp,Konoplya:2017wot,Konoplya:2022gjp,Bolokhov:2023dxq}. However, Dirac QNMs exhibit unique features, such as spin-orbit coupling effects, that distinguish them from scalar or gravitational perturbations. These modes also play a some role in the context of black hole thermodynamics and quantum gravity, where they can be used to probe the underlying quantum structure of spacetime. Analytical and numerical methods, including the WKB approximation and time-domain integration, are commonly employed to calculate these modes, revealing how different black hole parameters, such as mass, charge, and spin, influence the fermionic QNM spectrum. As a result, quasinormal modes of Dirac fields have been extensively studied for a great number of black hole models and gravitational theories
 \cite{Zinhailo:2019rwd,Bolokhov:2024ixe,Gonzalez:2014voa,Jing:2005uy,Konoplya:2017tvu,Rosa:2011my,Wongjun:2019ydo,Kanti:2006ua,Al-Badawi:2023xig,Chen:2005rm,Konoplya:2022hll,Saleh:2016pke,Chakrabarti:2008xz,Varghese:2010qv,Sebastian:2014dka,Lopez-Ortega:2007vlo,Wu:2004vb,Zhang:2005zs,Lopez-Ortega:2010lld,Wang:2009aj,Wang:2017fie,Konoplya:2007zx,Jing:2003wq,Al-Badawi:2022aby,Zhou:2003ts,Konoplya:2020jgt,Lopez-Ortega:2014daa}.

There exist many different approaches to constructing quantum-corrected black hole spacetimes. As a result, the literature on perturbations and quasinormal modes of such quantum-corrected black holes is vast. One line of research focuses on black holes obtained by accounting for spherically symmetric quantum fluctuations of the metric and matter fields in a nonperturbative manner~\cite{Kazakov:1993ha}, with their quasinormal spectra analyzed in~\cite{Saleh:2016pke,Saleh:2014uca,Konoplya:2019xmn}. Another major direction involves higher-curvature corrected theories, which are motivated by quantum corrections in the context of string theory. Black hole perturbations in such models have been studied in various dimensions: in \(2+1\) dimensions~\cite{Konoplya:2020ibi,Skvortsova:2023zca,Skvortsova:2023zmj}, four dimensions~\cite{Konoplya:2020bxa,Zinhailo:2018ska,Cano:2020cao,Wagle:2021tam,Konoplya:2019hml,Matyjasek:2020bzc,Blazquez-Salcedo:2020caw,Blazquez-Salcedo:2022omw,Konoplya:2022iyn}, and in higher-dimensional cases~\cite{Abdalla:2005hu,Gonzalez:2018xrq,Cuyubamba:2016cug,Yoshida:2015vua,Konoplya:2017ymp,Chen:2015fuf}.

A large body of recent work has also focused on quasinormal modes of black holes in the framework of asymptotically safe gravity~\cite{Stashko:2024wuq,Dubinsky:2024aeu,Malik:2024tuf,Konoplya:2023aph,Pedrotti:2024znu}. Additionally, various loop quantum gravity–inspired approaches, such as the Hamiltonian constraint method and holonomy corrections, have been employed to study the spectra of quantum-corrected black holes~\cite{Bolokhov:2023bwm,Bolokhov:2023ruj,Liu:2020ola,Skvortsova:2024atk,Konoplya:2025hgp,Yang:2024ofe,Gong:2023ghh,Moreira:2023cxy,Zinhailo:2024kbq,del-Corral:2022kbk}. Finally, the quasinormal spectrum of a quantum-corrected black hole derived from T-duality arguments has been investigated in~\cite{Nicolini:2019irw,Konoplya:2023ahd}.

One the promising approaches to construction of quantum gravity is the Loop Quantum Gravity \cite{Ashtekar:2004eh,Ashtekar:1987gu,Ashtekar:1996eg}, the test for which in the electromagnetic and gravitational sectors have been studied in \cite{Vagnozzi:2022moj,Afrin:2022ztr}.
Recently, within the Hamiltonian constraints approach, intriguing models for spherically symmetric quantum-corrected black holes that preserve general covariance have been developed in \cite{Zhang:2024khj}. As a result, various physical properties of these black holes, such as optical phenomena, accretion and particle motion, have been studied in \cite{Liu:2024soc,Liu:2024wal,Chen:2025ifv,Shu:2024tut}. The quasinormal modes of boson fields for these models have also been recently investigated in \cite{Konoplya:2024lch,Bolokhov:2024bke,Malik:2024elk}. In this work, we extend these studies to consider massless Dirac field perturbations, implying, first of all, propagating neutrinos in a black hole background, and calculate the corresponding quasinormal frequencies using two alternative methods: the higher-order WKB method with Padé approximants and time-domain integration. We will demonstrate that the quasinormal modes of the Dirac field exhibit a noticeable shift when the quantum parameter is introduced. In addition we obtain the compact analytic formula for quasinormal frequencies in the form of the expansion beyond the eikonal limit.
Using the correspondence between the grey-body factors and quasinormal modes \cite{Konoplya:2024lir} this formula gives an analytic expression for the grey-body factors for neutrinos.

This paper is organized as follows: In Sec. II, we introduce the black hole metrics, derive the wave equations, and discuss the corresponding effective potentials. In Sec. III, we review the methods used for the calculations. Sec. IV presents the calculated quasinormal frequencies, and the Conclusions summarize our results.

\section{Black hole metric and wave-like equation}\label{sec:wavelike}

The metric of the quantum corrected black hole is given by the following line element \cite{Zhang:2024khj}
\begin{equation}\label{metric}
  ds^2=-f(r)dt^2+\frac{B^2(r)}{f(r)}dr^2+r^2(d\theta^2+\sin^2\theta d\phi^2),
\end{equation}
where for the first model we have
\begin{equation}\label{metric1}
f(r)=\displaystyle 1  -\frac{2 M}{r} + \frac{\xi ^2 \left(1-\frac{2 M}{r}\right)^2}{r^2}, \quad B(r)=1, \quad first~model
\end{equation}
and for the second we have
\begin{equation}\label{metric2}
f(r)=\displaystyle 1-\frac{2 M}{r}, \quad B(r)=\displaystyle \frac{1}{\sqrt{\frac{\xi ^2 \left(1-\frac{2 M}{r}\right)}{r^2}+1}}, \quad second~model.
\end{equation}
Here $\xi $ is the coupling responsible for quantum corrections, and $M$ is the mass parameter.

The derivation of the general covariant Dirac equation in curved spacetime involves several key steps.
In curved spacetime, the metric $g_{\mu\nu}$ can be related to the local flat spacetime (Minkowski) metric $\eta_{ab}$ using the tetrads $e_\mu^a$ (also called vierbeins in 4D):
\[
g_{\mu\nu} = e_\mu^a e_\nu^b \eta_{ab},
\]
where $e_\mu^a$ are the components of the tetrad basis vectors. The inverse tetrads are defined by $e^\mu_a$, satisfying the relations:
\[
e_\mu^a e^\mu_b = \delta^a_b, \quad e_\mu^a e^\nu_a = \delta_\mu^\nu.
\]
The gamma matrices in curved spacetime $\gamma^\mu$ are related to the flat spacetime gamma matrices $\gamma^a$ through the tetrads:
\[
\gamma^\mu = e^\mu_a \gamma^a.
\]
These gamma matrices satisfy the Clifford algebra:
\[
\{\gamma^\mu, \gamma^\nu\} = 2 g^{\mu\nu} I,
\]
where $I$ is the identity matrix.
To account for the curvature of spacetime, we introduce the spin connection $\Gamma_\mu$. The spin connection is necessary to define a covariant derivative that acts on spinors. It is given by:
\[
\Gamma_\mu = \frac{1}{8} \omega_\mu^{ab} [\gamma_a, \gamma_b],
\]
where $\omega_\mu^{ab}$ is the spin connection 1-form, defined in terms of the tetrads:
\[
\omega_\mu^{ab} = e_\nu^a \nabla_\mu e^{\nu b},
\]
and $\nabla_\mu$ is the covariant derivative compatible with the metric $g_{\mu\nu}$.
The covariant derivative acting on the spinor field $\Psi$ is given by:
\[
D_\mu \Psi = \left( \partial_\mu - \Gamma_\mu \right) \Psi.
\]
This derivative ensures that the Dirac equation remains consistent under local Lorentz transformations and parallel transport in curved spacetime.
Finally, substituting the above components into the flat spacetime Dirac equation, we obtain the covariant Dirac equation in curved spacetime:
\[
i \gamma^\mu \left( \partial_\mu - \Gamma_\mu \right) \Psi =0 
\]
This equation describes the dynamics of spin-1/2 particles in a curved spacetime background, incorporating both gravitational effects and the intrinsic spin of the particles.

After separation of the variables in the background (\ref{metric}) the above equations take the Schrödinger wavelike form:
\begin{equation}\label{wave-equation}
\dfrac{d^2 \Psi}{dr_*^2}+(\omega^2-V(r))\Psi=0,
\end{equation}
where the ``tortoise coordinate'' $r_*$ is defined as follows:
\begin{equation}\label{tortoise}
dr_*\equiv\frac{B(r)}{f(r)}dr.
\end{equation}

For the Dirac field ($s=1/2$) one has two isospectral potentials,
\begin{equation}
V_{\pm}(r) = W^2\pm\frac{dW}{dr_*}, \quad W\equiv \left(\ell+\frac{1}{2}\right)\frac{\sqrt{f(r)}}{r}.
\end{equation}
The isospectral wave functions can be transformed one into another by the Darboux transformation,
\begin{equation}\label{psi}
\Psi_{+}\propto \left(W+\dfrac{d}{dr_*}\right) \Psi_{-},
\end{equation}
so that it is sufficient to calculate quasinormal modes for only one of the effective potentials. We will do that for $V_{+}(r)$ because the WKB method works better in this case.

\begin{figure}
\resizebox{\linewidth}{!}{\includegraphics{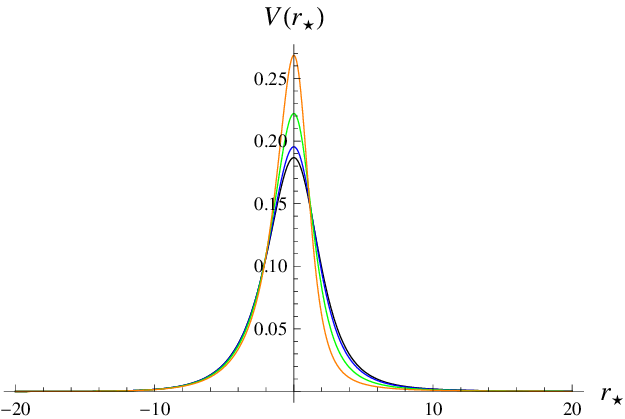}\includegraphics{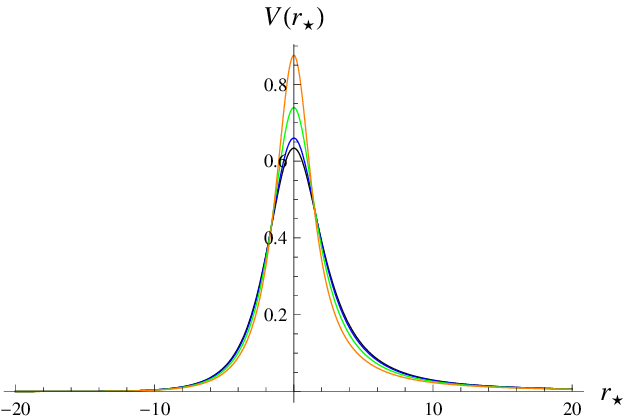}}
\caption{Potential as a function of the tortoise coordinate of the $\ell=1/2$ (left) and $\ell=3/2$ (right) for the first model of quantum corrected black hole ($M=1/2$): $\xi=0$ (black), $\xi=0.5$ (blue), $\xi=1$ (green), $\xi=1.5$ (orange).}\label{fig:potential1}
\end{figure}

\begin{figure}
\resizebox{\linewidth}{!}{\includegraphics{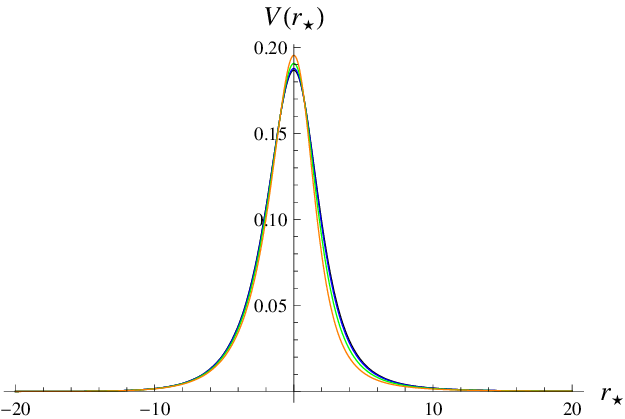}\includegraphics{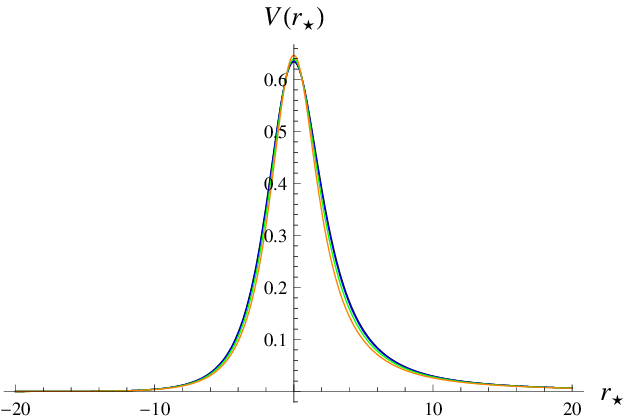}}
\caption{Potential as a function of the tortoise coordinate of the $\ell=1/2$ (left) and $\ell=3/2$ (right) for the second model of quantum corrected black hole ($M=1/2$): $\xi=0$ (black), $\xi=0.5$ (blue), $\xi=1$ (green), $\xi=1.5$ (orange).}\label{fig:potential2}
\end{figure}

Effective potentials for both black hole models and various values of $\xi$ are shown in figs. \ref{fig:potential1} and \ref{fig:potential2}. One the two iso-spectral effective potentials $V_{-}$  for the Dirac field has a negative gap near the event horizon. However, the other potential $V_{+}$ is positive definite, which provides stability of the perturbations.

\begin{table*}
\begin{tabular}{c c c c c}
\hline
\hline
$\xi $ & Prony fit & WKB6 Padé & rel. error $Re (\omega)$ & rel. error $Im (\omega)$  \\
\hline
$0$ & $0.36609-0.19401 i$ & $0.365286-0.193132 i$ & $0.219\%$ & $0.452\%$\\
$0.2$ & $0.36619-0.19433 i$ & $0.366551-0.194854 i$ & $0.0990\%$ & $0.268\%$\\
$0.4$ & $0.36649-0.19530 i$ & $0.368199-0.199011 i$ & $0.466\%$ & $1.90\%$\\
$0.6$ & $0.36698-0.19692 i$ & $0.371304-0.204026 i$ & $1.18\%$ & $3.61\%$\\
$0.8$ & $0.36766-0.19918 i$ & $0.373520-0.208761 i$ & $1.59\%$ & $4.81\%$\\
$1.$ & $0.36850-0.20209 i$ & $0.370579-0.217833 i$ & $0.564\%$ & $7.79\%$\\
$1.2$ & $0.36949-0.20563 i$ & $0.370030-0.234753 i$ & $0.147\%$ & $14.2\%$\\
\hline
\hline
\end{tabular}
\caption{QNMs for the first model by  time-domain integration and WKB methods: $s=1/2$, $\ell=1/2$, $M=1/2$.}\label{tableI}
\end{table*}

\begin{table*}
\begin{tabular}{c c c c c}
\hline
\hline
$\xi $ & Prony fit & WKB6 Padé & rel. error $Re (\omega)$ & rel. error $Im (\omega)$  \\
\hline
$0$ & $0.759199-0.191028 i$ & $0.760082-0.192816 i$ & $0.116\%$ & $0.936\%$\\
$0.2$ & $0.760956-0.192602 i$ & $0.761892-0.194390 i$ & $0.123\%$ & $0.928\%$\\
$0.4$ & $0.766208-0.197315 i$ & $0.767283-0.199141 i$ & $0.140\%$ & $0.925\%$\\
$0.6$ & $0.774902-0.205119 i$ & $0.776314-0.207030 i$ & $0.182\%$ & $0.931\%$\\
$0.8$ & $0.786957-0.215827 i$ & $0.788404-0.217918 i$ & $0.184\%$ & $0.969\%$\\
$1.$ & $0.801692-0.228755 i$ & $0.803530-0.232082 i$ & $0.229\%$ & $1.45\%$\\
$1.2$ & $0.817989-0.247162 i$ & $0.821433-0.249403 i$ & $0.421\%$ & $0.907\%$\\
\hline
\hline
\end{tabular}
\caption{QNMs for the first model by  time-domain integration and WKB methods:  $s=1/2$, $\ell=3/2$, $M=1/2$.}\label{tableII}
\end{table*}

\begin{table*}
\begin{tabular}{c c c c c}
\hline
\hline
$\xi $ & Prony fit & WKB6 Padé & rel. error $Re (\omega)$ & rel. error $Im (\omega)$  \\
\hline
$0$ & $1.148857-0.191519 i$ & $1.148189-0.192609 i$ & $0.0582\%$ & $0.569\%$\\
$0.2$ & $1.151879-0.193218 i$ & $1.151276-0.194255 i$ & $0.0524\%$ & $0.537\%$\\
$0.4$ & $1.160915-0.198314 i$ & $1.160483-0.199189 i$ & $0.0372\%$ & $0.441\%$\\
$0.6$ & $1.175865-0.206800 i$ & $1.175642-0.207403 i$ & $0.0190\%$ & $0.291\%$\\
$0.8$ & $1.196565-0.218654 i$ & $1.196489-0.218887 i$ & $0.00635\%$ & $0.106\%$\\
$1.$ & $1.222769-0.233828 i$ & $1.222677-0.233630 i$ & $0.00750\%$ & $0.0849\%$\\
$1.2$ & $1.254129-0.252283 i$ & $1.253824-0.251626 i$ & $0.0243\%$ & $0.260\%$\\
\hline
\hline
\end{tabular}
\caption{QNMs for the first model by  time-domain integration and WKB methods:  $s=1/2$, $\ell=5/2$, $M=1/2$.}\label{tableIII}
\end{table*}

\begin{table*}
\begin{tabular}{c c c c c}
\hline
\hline
$\xi $ & Prony fit & WKB6 Padé & rel. error $Re (\omega)$ & rel. error $Im (\omega)$  \\
\hline
$0$ & $0.375023-0.185524 i$ & $0.365286-0.193132 i$ & $2.60\%$ & $4.10\%$\\
$0.2$ & $0.374590-0.186107 i$ & $0.365513-0.193619 i$ & $2.42\%$ & $4.04\%$\\
$0.4$ & $0.373341-0.187836 i$ & $0.365010-0.195645 i$ & $2.23\%$ & $4.16\%$\\
$0.6$ & $0.371392-0.190633 i$ & $0.363543-0.197994 i$ & $2.11\%$ & $3.86\%$\\
$0.8$ & $0.368836-0.194359 i$ & $0.361813-0.200890 i$ & $1.90\%$ & $3.36\%$\\
$1.$ & $0.365708-0.198847 i$ & $0.359952-0.204471 i$ & $1.57\%$ & $2.83\%$\\
$1.2$ & $0.362003-0.203928 i$ & $0.358314-0.208717 i$ & $1.02\%$ & $2.35\%$\\
\hline
\hline
\end{tabular}
\caption{QNMs for the second model by  time-domain integration and WKB methods: $s=1/2$, $\ell=1/2$, $M=1/2$.}\label{tableIV}
\end{table*}

\begin{table*}
\begin{tabular}{c c c c c}
\hline
\hline
$\xi $ & Prony fit & WKB6 Padé & rel. error $Re (\omega)$ & rel. error $Im (\omega)$  \\
\hline
$0$ & $0.759199-0.191028 i$ & $0.760082-0.192816 i$ & $0.116\%$ & $0.936\%$\\
$0.2$ & $0.759006-0.191541 i$ & $0.759899-0.193333 i$ & $0.118\%$ & $0.936\%$\\
$0.4$ & $0.758431-0.193070 i$ & $0.759352-0.194880 i$ & $0.121\%$ & $0.938\%$\\
$0.6$ & $0.757482-0.195582 i$ & $0.758438-0.197430 i$ & $0.126\%$ & $0.945\%$\\
$0.8$ & $0.756173-0.199027 i$ & $0.757161-0.200963 i$ & $0.131\%$ & $0.973\%$\\
$1.$ & $0.754519-0.203339 i$ & $0.755631-0.205476 i$ & $0.147\%$ & $1.05\%$\\
$1.2$ & $0.752536-0.208437 i$ & $0.753891-0.210536 i$ & $0.180\%$ & $1.01\%$\\
\hline
\hline
\end{tabular}
\caption{QNMs for the second model by  time-domain integration and WKB methods: $s=1/2$, $\ell=3/2$, $M=1/2$.}\label{tableV}
\end{table*}

\begin{table*}
\begin{tabular}{c c c c c}
\hline
\hline
$\xi $ & Prony fit & WKB6 Padé & rel. error $Re (\omega)$ & rel. error $Im (\omega)$  \\
\hline
$0$ & $1.148857-0.191519 i$ & $1.148189-0.192609 i$ & $0.0582\%$ & $0.569\%$\\
$0.2$ & $1.148702-0.192095 i$ & $1.148060-0.193155 i$ & $0.0559\%$ & $0.552\%$\\
$0.4$ & $1.148243-0.193808 i$ & $1.147676-0.194780 i$ & $0.0494\%$ & $0.501\%$\\
$0.6$ & $1.147487-0.196621 i$ & $1.147035-0.197455 i$ & $0.0395\%$ & $0.424\%$\\
$0.8$ & $1.146450-0.200474 i$ & $1.146138-0.201131 i$ & $0.0272\%$ & $0.328\%$\\
$1.$ & $1.145145-0.205294 i$ & $1.144987-0.205749 i$ & $0.0138\%$ & $0.222\%$\\
$1.2$ & $1.143585-0.210997 i$ & $1.143584-0.211238 i$ & $0.00011\%$ & $0.114\%$\\
\hline
\hline
\end{tabular}
\caption{QNMs for the first model by  time-domain integration and WKB methods: $s=1/2$, $\ell=5/2$, $M=1/2$.}\label{tableVI}
\end{table*}

\section{WKB method and time-domain integration}\label{sec:methods}

\subsection{WKB Method}

For a given effective potential and metric  the quasinormal modes satisfy the following boundary conditions:
\begin{equation}\label{boundaryconditions}
\Psi(r_*\to\pm\infty)\propto e^{\pm\imo \omega r_*},
\end{equation}
which ensure that the wave is purely ingoing at the event horizon ($r_*\to-\infty$) and purely outgoing at spatial infinity ($r_*\to\infty$).

If the potential can be expressed as \( V(r, \omega) = V(r) - \omega^2 \), where the frequency \( \omega \) is separated from the effective potential \( V(r) \), the WKB quantization condition for quasinormal modes (QNMs) can be written as:
\[
\omega_n^2 = V_0 - \left( \frac{n + \frac{1}{2}}{\sqrt{2}} \sqrt{2V_0''} + \sum_{j=1}^{\infty} \Lambda_j \right)
\]
This expression provides the square of the QNM frequencies \( \omega_n \) in terms of the maximum of the potential \( V_0 \), its second derivative \( V_0'' \), and the higher-order WKB corrections \( \Lambda_j \). The sum of higher-order corrections \( \sum_{j=1}^{\infty} \Lambda_j \) ensures the accuracy of the QNM frequency calculation, especially when going beyond the leading WKB order.
The explicit forms of the corrections have been derived  in \cite{Iyer:1986np} up to the third order,  and in \cite{Konoplya:2003ii} and \cite{Matyjasek:2017psv} up to the sixth and 13th orders respectively. The WKB method has been used for finding quasinormal modes in numerous works (see, for example, \cite{Konoplya:2006ar,Balart:2023odm,Breton:2017hwe,Yu:2022yyv,Konoplya:2022hbl,Al-Badawi:2023lvx,Xiong:2021cth,Kokkotas:2010zd}), so that we will not discuss it here in more detail.
Here we will use the 6th order WKB method \cite{Konoplya:2003ii}  with Padé approximants \cite{Matyjasek:2017psv,Konoplya:2019hlu}.

\subsection{Time-Domain Integration}

To verify the accuracy of the WKB method, we employ time-domain integration as an independent approach. For this purpose, we use the Gundlach-Price-Pullin discretization scheme \cite{Gundlach:1993tp}. The time-domain integration is based on the numerical evolution of the wave equation in double-null coordinates $u=t-r_*$ and $v=t+r_*$, where $r_*$ is the tortoise coordinate.

The discretized wave equation in the time domain can be expressed as:
\begin{equation}
\Psi(N) = \Psi(W) + \Psi(E) - \Psi(S)   - \Delta^2 V(S) \frac{\Psi(W) + \Psi(E)}{8} + \mathcal{O}\left(\Delta^4\right), \label{Discretization}
\end{equation}
where the points $N$, $W$, $E$, and $S$ are defined as:
\begin{eqnarray}
N &\equiv& \left(u+\Delta,v+\Delta\right), \\
W &\equiv& \left(u+\Delta,v\right), \\
E &\equiv& \left(u,v+\Delta\right), \\
S &\equiv& \left(u,v\right).
\end{eqnarray}

This scheme is derived by expanding the wave function $\Psi$ in a Taylor series at these grid points and matching terms up to the required order of accuracy. The advantage of this approach is that it provides a stable and convergent numerical solution for the wave equation, capturing the full evolution of the perturbation from initial data through the ringdown phase. This method has been widely used and has proven to be highly accurate in numerous works \cite{Churilova:2021tgn,Momennia:2022tug,Qian:2022kaq,Churilova:2019qph,Abdalla:2012si}.

The Prony method is a mathematical technique used for fitting a sum of exponentials to a given set of data points, particularly useful in signal processing and system identification. It models data as a sum of damped exponentials, expressed as:
\[
\Psi(t) = \sum_{k=1}^{N} A_k e^{j\omega_k t}
\]
where \( A_k \) represents the amplitude, \( \alpha_k \) is the decay rate, \( \omega_k \) is the frequency, \( t \) is the time, and \( j \) is the imaginary unit. The method involves solving a set of linear equations derived from the data, often reducing to an eigenvalue problem to determine these parameters. While powerful, the method can be sensitive to noise and requires uniformly sampled data.

\section{Quasinormal modes}\label{sec:QNM}

Tables I-VI present the values of quasinormal modes calculated using the 6th-order WKB method and time-domain integration. It is evident that for \(\ell = 1/2\), the discrepancy between the two methods is quite large and comparable to the effect resulting from the deviation from Schwarzschild geometry.

Given that the WKB method converges only asymptotically, we rely on the time-domain integration for the \(\ell = 1/2\) case. However, for higher \(\ell\) values, the discrepancy is already at least one order of magnitude smaller than the effect itself.

For the second black hole model, we observe that the oscillation frequency \( \text{Re}(\omega) \) decreases as \(\xi\) increases, while the damping rate, proportional to \(\text{Im}(\omega)\), increases with growing \(\xi\). This indicates that the quality factor of the oscillations decreases with \(\xi\), meaning that the quantum-corrected black hole radiates fermions less effectively when quantum corrections are considered, according to the second model. In contrast, for the first black hole model, both the oscillation frequency and the damping rate increase with \(\xi\).

Here, we will also derive an analytic approximation for quasinormal modes using an expansion in powers of the inverse multipole number, as suggested in \cite{Konoplya:2023moy} and applied in various studies for different black hole models \cite{Bolokhov:2024ixe,Malik:2024tuf,Malik:2024sxv,Malik:2024voy,Malik:2023bxc} and reviewed in \cite{Bolokhov:2025uxz}.
Perturbations in a spherically symmetric background can be reduced to the wave-like equation with the effective potential which can be approximated in the following way:
\begin{equation}\label{potential-multipole}
V(r_*)=\kappa^2\left(H_{0}(r_*)+H_{1}(r_*) \kappa^{-1} +H_{2}(r_*) \kappa^{-2}+\ldots\right).
\end{equation}
Here $\kappa\equiv\ell+\frac{1}{2}$ and $\ell=s,s+1,s+2,\ldots$ is the positive half(integer) multipole number, which has minimal value equal to the spin of the field under consideration $s$. Here, following \cite{Konoplya:2023moy} we use an expansion in terms of $\kappa^{-1}$.

The function $H(r_*)$ has a single peak, so that, the location of the potential's maximum (\ref{potential-multipole}) can be expanded as follows
\begin{equation}\label{rmax}
  r_{\max }=r_0+r_1\kappa^{-1}+r_2\kappa^{-2}+\ldots.
\end{equation}

Substituting (\ref{rmax}) into the following first order WKB formula
\begin{eqnarray}
\omega&=&\sqrt{V_0-\imo \K\sqrt{-2V_2}},
\end{eqnarray}
and then expanding in $\kappa^{-1}$, we find that,
\begin{eqnarray}\label{eikonal-formulas}
\omega=\Omega\kappa-\imo\lambda\K+\Order{\kappa^{-1}}.
\end{eqnarray}
The above relation is a reasonable approximation for $\kappa\gg\K$.

Then, for the first black hole model for the location of the maximum we have
\begin{widetext}
\begin{equation}\label{rmax-Dirac}
r_{\max } = \frac{11 M}{16 \sqrt{3} \kappa ^3}-\frac{\sqrt{3} M}{2 \kappa }+3 M+\xi^2 \left(\frac{11}{96 \sqrt{3} M \kappa ^3}-\frac{1}{36 \sqrt{3} M \kappa }\right)+\mathcal{O}\left(\xi^4,\frac{1}{\kappa ^4}\right)
\end{equation}
and the quasinormal modes are
\begin{equation}\label{eikonal-Dirac}
\begin{array}{rcl}
\omega  &=& \displaystyle\frac{i \K \left(119-940 \K^2\right)}{46656 \sqrt{3} M \kappa ^2}-\frac{60 \K^2+7}{1296 \sqrt{3} M \kappa }+\frac{\kappa }{3 \sqrt{3} M}-\frac{i \K}{3 \sqrt{3} M}\\
&&\displaystyle+\xi^2 \left(-\frac{55 i \K \left(4 \K^2-101\right)}{839808 \sqrt{3} M^3 \kappa ^2}-\frac{276 \K^2+61}{23328 \sqrt{3} M^3 \kappa }+\frac{\kappa }{162 \sqrt{3} M^3}-\frac{i \K}{54 \sqrt{3} M^3}\right)+\mathcal{O}\left(\xi^4,\frac{1}{\kappa ^3}\right).
\end{array}
\end{equation}
\end{widetext}

For the second black hole model we have:
\begin{equation}\label{rmax-Dirac}
r_{\max } = \frac{11 M}{16 \sqrt{3} \kappa ^3}-\frac{\sqrt{3} M}{2 \kappa }+3 M+\xi^2 \left(\frac{17}{288 \sqrt{3} M \kappa ^3}-\frac{1}{36 \sqrt{3} M \kappa }\right)+\mathcal{O}\left(\xi^4,\frac{1}{\kappa ^4}\right)
\end{equation}
and the quasinormal modes
\begin{equation}\label{eikonal-Dirac}
\begin{array}{rcl}
\omega  &=& \displaystyle\frac{i \K \left(119-940 \K^2\right)}{46656 \sqrt{3} M \kappa ^2}-\frac{60 K^2+7}{1296 \sqrt{3} M \kappa }+\frac{\kappa }{3 \sqrt{3} M}-\frac{i \K}{3 \sqrt{3} M}\\
&&\displaystyle+\xi^2 \left(\frac{i \left(2099-220 \K^2\right) \K}{839808 \sqrt{3} M^3 \kappa ^2}-\frac{84 K^2+17}{17496 \sqrt{3} M^3 \kappa }-\frac{i \K}{162 \sqrt{3} M^3}\right)+\mathcal{O}\left(\xi^4,\frac{1}{\kappa ^3}\right)
\end{array}
\end{equation}

In \cite{Cardoso:2008bp}, it was demonstrated that there is a correspondence between the parameters of null geodesics (orbital frequency and Lyapunov exponent) and the real and imaginary parts of quasinormal modes in the eikonal limit. This correspondence is known to hold in the majority of cases. However, there are several exceptions and limitations, as discussed in \cite{Konoplya:2017wot,Konoplya:2022gjp,Bolokhov:2023dxq}.  The eikonal limit is interesting not only because in this regime the frequencies could be found in analytic form (see, for instance,  \cite{Konoplya:2005sy,Bolokhov:2023bwm,Konoplya:2001ji,Zhidenko:2003wq,Allahyari:2018cmg}), but also because it may bring a catastrophic instability \cite{Konoplya:2017lhs,Takahashi:2011qda,Konoplya:2017zwo,Gleiser:2005ra,Dotti:2005sq} in theories with higher corrections in curvature.

For the present case of Dirac field perturbations, in the eikonal limit perturbation equations do not depend on the spin of the field and we reproduce eqs. (33) and (34) of \cite{Konoplya:2024lch}, and, consequently, confirm the correspondence. In \cite{Konoplya:2024lch}, the existence of non-perturbative quasinormal modes in the parameter \(\xi\) was reported, which apparently cannot be reproduced by the WKB method and, therefore, do not satisfy the correspondence. However, such modes appear in the spectrum at sufficiently large values of \(\xi\), while the underlying theory itself is perturbative in \(\xi\). Thus, we conclude that the emergence of non-perturbative modes likely indicates a breakdown of the perturbative regime of quantum corrections, rather than a breakdown of the correspondence.

It is worth noting that the analytically derived expressions above allow one to compute not only the fundamental quasinormal mode but also overtones, via their dependence on the parameter \( K = n + \tfrac{1}{2} \). However, the WKB approximation remains sufficiently accurate only in the regime where the overtone number \( n \) is smaller than the multipole number \( \ell \). Consequently, it cannot capture the so-called ``outburst of overtones'' observed in~\cite{Konoplya:2022pbc}, which refers to a strong deviation of overtone frequencies from their Schwarzschild values as \( n \) increases. Although studies of integer-spin perturbations in quantum-corrected black holes~\cite{Konoplya:2024lch} suggest the presence of significant deviations in overtone spectra, such overtones are not typically visible in time-domain profiles. This is because the fundamental mode dominates the energy content of the ringdown signal. Indeed, we were also unable to extract overtones from time-domain profiles in the case of Dirac perturbations. However, using an alternative representation of the Dirac wave equation—one that avoids square roots—and subsequently applying the precise Leaver method \cite{Leaver:1985ax}, would allow accurate extraction of overtone frequencies even in the regime \( \ell \leq n \).

In addition to the quasinormal modes boundary conditions problem, another important spectral problem is related to the Hawking radiation of quantum corrected black holes, which has been actively studied recently in \cite{Calza:2024xdh,Calza:2024fzo}. This is the scattering problem which allows one to find the portions of reflected and transmitted particles of the initial Hawking radiation flow and thereby to find the total Hawking evaporation rate.

The obtained here analytic expression for quasinormal modes can be effective used for finding aanalytic approximation for the grey-body factors through the correcspondence between quasinormal modes and grey-body factors \cite{Konoplya:2024lir}. Thus taking the fundamental mode $\omega_{0 \ell}$ and the first overtone
$\omega_{1 \ell}$ for a given multipole number $\ell$ and using eq. 4.6 of \cite{Konoplya:2024lir}, we can find the grey-body factors as follows:
\begin{eqnarray}\label{transmission-eikonal}
\Gamma_{\ell}(\Omega)\equiv|T|^2&=&\left(1+e^{2\pi\dfrac{\Omega^2-\re{\omega_{0 \ell}}^2}{4\re{\omega_{0 \ell}}\im{\omega_{0 \ell}}}}\right)^{-1} + beyond~eikonal~corrections.
\end{eqnarray}
Here $\Omega$ is the real frequency of the wave in the scattering problems and the beyond-eikonal corrections (which depend also on $\omega_{1 \ell}$)  can be found in eqs. (4.2-4.5) of \cite{Konoplya:2024lir}. The above equation together with the obtained here values $\omega_{0}$ and $\omega_{1}$ gives the transimission coefficient for neutrinos in the background of the considered quantum corrected black holes.

\section{Conclusions}\label{sec:conclusions}

Recent work \cite{Zhang:2024khj} derives and analyzes quantum-modified black hole spacetimes that address the issue of general covariance in spherically symmetric gravity by applying minimal requirements to the effective Hamiltonian constraint. While the quasinormal modes of bosonic fields have been thoroughly studied in recent works \cite{Konoplya:2024lch,Bolokhov:2024bke,Malik:2024elk}, no such analysis has been conducted for fermionic perturbations. In this paper, we complement previous studies by calculating the quasinormal modes of the massless Dirac field, implying propagation of neutrinos, in the background of the two models of quantum-corrected black holes developed in \cite{Zhang:2024khj}. The data obtained through time-domain integration, with complex frequencies extracted using the Prony method, is consistent with the results from the 6th-order WKB method with Padé approximants within the range of parameters where both methods are reliable. Additionally, we derive an analytic formula for the frequencies as an expansion beyond the eikonal limit. The two models of quantum-corrected black holes can be distinguished by their quasinormal spectra. The obtained analytic formula for quasinormal modes can also be used to obtain the analytic expression for the neutrino's grey-body factors. Our work could be extended by considering massive Dirac perturbations, which, as massive field of every spin, exhibit several qualitative differences both during the ringdown phase and in the asymptotic tail behavior~\cite{Ohashi:2004wr,Konoplya:2006gq,Bolokhov:2023bwm,Koyama:2000hj,Lutfuoglu:2025hjy}.

\begin{acknowledgments}
The author thanks R. A. Konoplya for useful discussions.
\end{acknowledgments}

\bibliography{bibliographyWW}
\end{document}